\documentclass{jfm}

\usepackage{natbib}
\usepackage{graphicx,subfigure}

\ifCUPmtlplainloaded \else
  \checkfont{eurm10}
  \iffontfound
    \IfFileExists{upmath.sty}
      {\typeout{^^JFound AMS Euler Roman fonts on the system,
                   using the 'upmath' package.^^J}%
       \usepackage{upmath}}
      {\typeout{^^JFound AMS Euler Roman fonts on the system, but you
                   dont seem to have the}%
       \typeout{'upmath' package installed. JFM.cls can take advantage
                 of these fonts,^^Jif you use 'upmath' package.^^J}%
      }
  \else
  \fi
\fi

\ifCUPmtlplainloaded \else
  \checkfont{msam10}
  \iffontfound
    \IfFileExists{amssymb.sty}
      {\typeout{^^JFound AMS Symbol fonts on the system, using the
                'amssymb' package.^^J}%
       \usepackage{amssymb}%

      }{}
  \fi
\fi

\ifCUPmtlplainloaded \else
  \IfFileExists{amsbsy.sty}
    {\typeout{^^JFound the 'amsbsy' package on the system, using it.^^J}%
     \usepackage{amsbsy}}
    {\providecommand\boldsymbol[1]{\mbox{\boldmath $##1$}}}
\fi

\newcommand\oneovers{\nobreak\mbox{$\;$s$^{-1}$}}
\newcommand\mpers{\nobreak\mbox{$\;$m\,s$^{-1}$}}
\newcommand\m{\nobreak\mbox{$\;$m${}$}}
\newcommand\km{\nobreak\mbox{$\;$km${}$}}
\newcommand\oneoverm{\nobreak\mbox{$\;$m$^{-1}$}}

\newcommand\F{\mbox{\textit{F}}}            
\newsavebox{\astrutbox}
\sbox{\astrutbox}{\rule[-5pt]{0pt}{20pt}}

\newcommand\p{\ensuremath{\partial}}

\newcommand\Gone{\ensuremath{\widetilde{G_{1}}}}
\newcommand\Gtwo{\ensuremath{\widetilde{G_{2}}}}%

\title [Non-Hydrostatic Effects on the Interaction between Flows and Orography]{Non-Hydrostatic Effects in the Interaction between Flow and Orography}

\author[I. Gladich, F.  Stel, D.  Giaiotti and G. Furlan]%
{I\ls V\ls A\ls N\ns  G\ls L\ls A\ls D\ls I\ls C\ls H$^1$, \ns
F.\ls  S\ls T\ls E\ls  L$^2$,\break
\ns D.\ls  G\ls I\ls A\ls I\ls O\ls  T\ls T\ls   I$^2$ \and G.\ns F\ls U\ls R\ls L\ls A\ls N$^3$}

\affiliation{$^1$ Department of Mathematics, University of Trieste, Italy\\[\affilskip]
$^2$ ARPA-CRMA, via Cairoli 14, I-33057, Palmanova (UD), 
Italy \\[\affilskip]
$^3$the Abdus Salam
International Centre for Theoretical Physics (ICTP) Strada Costiera, 11
I-34014 Trieste
Italy}

\pubyear{1996}
\volume{538}
\pagerange{119--126}
\date{?? and in revised form ??}

\begin{document}

\maketitle

\begin{abstract}
The interaction between flows and orography is a fundamental aspect of theoretical 
fluid dynamics for its direct applications (e.g., in dynamical
meteorology); a comprehensive description is nowadays still lacking in
some aspects. In this work, in particular, the authors would like to
face the problem of flow-blocking and of the streamlines pattern
formation, examining the role of stratification (i.e., Brunt-Vaisala
frequency) and Froude number on these problems. In
particular  this work wants to investigate the role of vertical
advection on flow-blocking and on streamlines geometry. The importance
of streamlines curvature and stratification for the formation of
pressure perturbation, then their role in flow-blocking will be
shown. Moreover it will be shown how flow-blocking cannot be easly
predict using only a stratification parameter or the Froude number. 
\end{abstract}

\section{Introduction}


The interaction beetween flow and orography is an important topic of theoretical 
fluid dynamics because of its direct applications in everyday life. As an example, 
orographic rain is originated by a moist flow that, interacting with orography, gives rise to a vertical motion, then to condensation and precipitation formation.
Even if these phenomena are very common, their explanation is nowadays not complete. Infact a full description requires the knowleadge of the solution of Navier-Stokes equation with complex boundary conditions (i.e. top of troposphere and the orography).

The literature facing the interactions between flows and orography  can
be divided in: {\it i)} numerical works, {\it ii)} analytical works and {\it iii)}
experimental  works: \cite{Riley1976}, \cite{Baines1979}, \cite{Hunt1980},
  \cite{Castro1983} and   \cite{Snyder1985}. 
Among the numerical studies there are several contributions produced
using hydrostatic numerical models  \cite[]{Smolarkiewcz1989} and
more recently some contributions realized using non-hydrostatic
numerical models (e.g., the Weather, Research and Forecasting model,
WRF) \cite[] {Miglietta2005}. The problem  with numerical models,
both hydrostatic and non-hydrostatic, is that their output is
extremely complex then, generally,  very difficult to interpret
weighting the physical role of every possible parameter used in the
tuning of the model  \cite[]{Giaiotti2007b}. 
On the contrary analytic works permit to keep a more complete control
of the role of each parameter inserted into the analytical model even
if some approximations need to be taken to reduce the mathematical
difficulty of the starting equations. 

Concerning the analytical works, several of them  make use of the hydrostatic approximation and of
different kind of obstacles \cite[]{Lilly1979}, using both stratified and
rotational fluids \cite[]{McInttyre1972},  using thermal forcing
\cite[]{Reisner1993} or imposing turbulent  boundary layer at the
surface \cite[]{Carrunthers1990}. Only a few analitical works avoid the
use of the hydrostatic approximation, this because the vertical advective
term makes the analytic approch more difficult. A comprehensive review
of all these works can be found in \cite{Baines1995b}. 

Among the analytical works, three of them, \cite{Smith1989},
\cite{Wurtele1987} and \cite{Keller1994} deserve a special
mention. In particular \cite{Smith1989},  developing the previous work of \cite{Smith1988} and  \cite{Smith1989b}, studies the interaction between
hydrostatic and stratified flow on an idealized 3-D topography. In his
contribution the attention is focused on the case in which the flow stops its
upward motion while moving on the topography (i.e., stagnation of the
flow). Moreover, in this work, the use of Froude number as a discriminating
factor between stagnation and non-stagnation, proposed by \cite{Sheppard1956},
is critically reviewed. 
In the work of \cite{Keller1994} the study of interaction between non-hydrostatic and stratified flows on an idealized 2-D topography is presented but, in this case, the attention is focused on the effects of non-hydrostaticity on the formation of downstream lee waves. \cite{Keller1994} also analizes the behaviour of the flow with different vertical velocity profiles. The interesting aspect of \cite{Wurtele1987}, instead, stays in the approach to the gravity wave propagation in stratosphere.

Following the line defined by the three analytical works  above introduced, the aim of this work is to study analytically the influence of non-hydrostatic effects on the geometry of streamlines, essentially upstream to topography, and on flow blocking.
In the  ~\S\,\ref{sec:Hydro} a briefly review of \cite{Smith1989} is presented and its results are extended to the non-hydrostatic case. Then in
~\S\,\ref{sec:nohydro}, an analytical model for a stratified flow on a
2-D profile is presented adopting a uniform incident velocity profile
for the unperturbed flow and using the Fourier trasforms. In this part
the mathematical difficulty of antitrasformation is evidenced and a
new approach to overtake it is presented. With this approach 
the integrand is substituted with a new one for which the integration can be easly
carried out. 
In this way, making use of the new integral form, the streamlines
pattern for the non-hydrostatic case can be obtained and it is presented in 
~\S\,\ref{sec:stream}. With this approach the formation of lee waves,
already reproduced by \cite{Keller1994}, as well as the
intensification of wind speed at the top of topography profile is well
described by the model. This
phenomenon can be obtained even making use of hydrostatic models,  but
the non-hydrostatic approach showes a  minor intensification of
wind. Furthermore the changes in the pattern of streamline due to the
super-critical (Froude number $\F>1$) and sub-critical ($\F<1$)
regimes are presented. At the end of this work the relevance of
non-hydrostatic effects are shown to be important in the dynamics of
flow-blocking, ~\S\,\ref{sec:stagn}, specially for topographic
profiles with horizontal scale comparable with the vertical
one. Moreover, thanks to this study, stagnation can be put in
relationship with the formation of vorticity as guessed before by
\cite{Schar1993b} and \cite{Schar1993c}. At the end it is demonstrated
that simple parameters (e.g., Froude number or Brunt-Vaisala
frequency) are not sufficient to describe completely the stagnation
mechanism. This result is not merely theorethical but there are some
concrete cases in which it might had played a role, as is the case of
the Valcanale flood (29 August 2003, Valcanale-UD-, Italy) when two people
died.

\section{From hydrostatic to non-hydrostatic approach}\label{sec:Hydro}
In this section the work of Smith (1989) is adapted to the non-hydrostatic case.
A 2-D steady and parallel flow with a constant vertical velocity
profile is assumed. This flow interacts with an obstacle whose
analytic form is $ h=h(x,y)$. Differently from \cite{Smith1989} a 
bounded flow is here considered. Infact, imposing that all the
peturbation on pressure field and streamline displacement due to
orography damp at infinite height can create problems to the energy
conservation \cite[]{Baines1995b}. For this reason, in a more
realistic way, we impose that all the orographic effects fade at the
top of fluid that, in the atmospheric case coincides with the end of
troposphere  at $10$\km.

As done by \cite{Smith1989}, the flow is here assumed always parallel
(i.e., turbulent diffusion is neglected),  incompressible and stable
stratified with constant Brunt-Vaisala frequency $N$. In this
way far from orography the density profile $\rho_{\infty}(z)$ is horizontally homogeneous
and given by the following relationship 
\begin{equation}\label{eq:ro-inf}
\rho_{\infty}(z)=\rho_{0}\Big(1-\frac{N^{2}(z-z_{0})}{g}\Big)
\end{equation}
where $z$ is the vertical coordinate, $z_{0}$ is a reference level,
  $\rho_{0}$  the density at the reference level far
from the obstacles and $g$ is the gravity acceleration.

The flow is assumed as composed by dry air in isothermal condition and
all the sources and sinks of heat are neglected. The viscous effects
and the turbulent diffusion of momentum are neglected as
well. Moreover Coriolis force is assumed as null, infact Rossby number
is lower than unity for an incident velocity profile of $15$\mpers and
for a horizontal scale smaller than $200$\km. 

To proceed further it is now necessary to introduce an energy
conservation principle. For isentropic flows the energy for unit
volume $E$ given by
\begin{equation}\label{eq:energy}
E=\frac{\rho U^{2}}{2} + {P} +  \rho g z
\end{equation}
is constant along stream and vortex lines
\cite[]{Batchelor1994b}. In (\ref{eq:energy}) $P$ is the pressure
  and  $ U $ is the intensity of velocity vector. In
this case streamlines coincides  with isopicnal lines because the flow
is incompressible and steady. For this reason the energy balance
written for a parcel on a streamline characterized by the unperturbed
level $z_{0}$ (i.e., far from the obstacle) is conserved and has the
form 
 
$$
\frac{\rho_{0}{U^{2}}_{0}}{2} + \rho_{0}gz_{0} + P_{0}
 = \frac{\rho_{0}U^{2}}{2} + \rho_{0}gz + P 
$$

The pressure field far from orography is horizontally homogeneous as
previously assumed, then it is possible to assume the hydrostatic balance to describe $ P_{\infty}$
\begin{equation}\label{eq:energy}
\frac{\p{P_{\infty}(z)}}{\p{z}}=-\rho_{\infty}(z)g
\end{equation}
The reasonableness of this assumption  is given by the fact that
  no vertical motion, upstream and far from the obstacle, is assumed.
Defining the perturbation pressure $P^{*}$  as the difference between pressure field and pressure field at the same level and  far from the obstacles
$$
 P^{*}=P - P_{\infty}
$$
and using the equation (\ref{eq:energy}), the velocity of a parcel along a streamline is given by 
\begin{equation}\label{eq:bernoulli} 
U^{2} = {U^{2}}_{0} - N^{2}{\eta}^{2} -2\frac{P^{*}}{\rho_{0}}
\end{equation}
where ${U^{2}}_{0}$ is the velocity of an incident upstream vertical
profile while $\eta$ is the streamline vertical displacement relative
to the unperturbated streamline characterized by the height $z_{0}$
far from the obstacle and density $\rho_{0}$. 

The equation (\ref{eq:bernoulli}) gives us a simple relationship to identify the flow blocking, i.e. the situation in which $ U^{2} = 0 $. Neglecting the perturbation pressure term, the results of \cite{Sheppard1956} are reproduced. In those results stagnation occours when the displacement is
 
$$
\F=\frac{U}{N \eta} <1
$$

The problem of this approach to flow-blocking is that neglecting
pressure perturbation, the parcel behaves as if it were not immersed
in a fluid environment. It is then important, for a more realistic
description of the flow, try to estimate the pressure term. Starting from the vertical momentum equation \cite[]{Emanuel1994b} for a stratified flow
\begin{equation}\label{eq:dwdt}
\frac{Dw}{Dt}=-\frac{1}{\rho_{\infty}}\frac{\p{P^{*}}}{\p{z}}-\Big(\frac{\rho{'}}{\rho_{\infty}}\Big)g
\end{equation}
where $\rho'$ is the density  perturbation, i.e. the difference
  between density field and density field at the same level and far
  from the obstacles, that is
$$
\rho'=\rho(x,y,z)- \rho_{\infty}(z)=\frac{\rho_{0}N^{2}\eta}{g}
$$

The result of integration of (\ref{eq:dwdt}) from the general level
$z$ up to the top of the fluid $D$ becomes
$$
P^{*}(x,y,z,t) =\Gamma_{h}+
\Gamma_{nh}
$$
where
\begin{equation}\label{eq:ivan1}
\Gamma_{h} =  g\int_{z}^{D}\rho' dz  
\end{equation}
and
\begin{equation}\label{eq:gammahydro}
\Gamma_{nh} = \int_{z}^{D}\rho_{\infty}\frac{dw}{dt} dz
\end{equation}
that corrispond respectively to hydrostatic and non-hydrostatic
 contribution on the perturbation pressure

Using the isopicnal change of  coordinates $z = z_{0}+\eta $ in the first
integral $\Gamma_{h}$, equation (\ref{eq:energy}) it becomes
\begin{equation}\label{eq:U2}
U^{2} = {U^{2}}_{0} - 2N^{2}I_{\eta} -2\frac{ \Gamma_{nh}}{\rho_{0}}
\end{equation}
where
\begin{equation}\label{eq:Ieta}
I_{\eta}= \int_{z_{0}}^{D}\eta dz_{0}
\end{equation}
is the integral,  obtained integrating streamlines displacement $\eta$ from the unperturbed streamline
level $z_{0}$ to the top of the fluid where all streamlines
perturbation are null.

In \cite{Smith1989} the non-hydrostatic contribution $\Gamma_{nh}$ is
neglected. This means that in equation (\ref{eq:dwdt}) the vertical
advective term is neglected. Using scale analysis and
incompressibility condition it can be shown that the hydrostatic
assumption is good only for obstacles with a vertical scale lenght
smaller than the horizontal scale lenght. Using this approximation
\cite{Smith1989} showed that stagnation occurs when the condition 
\begin{equation}\label{eq:Fsmith}
\F=\frac{{U^{2}}_{0}} {2N^{2}I_{\eta}}<1
\end{equation}
is satisfied.  Stagnation dynamics is then connected to a non local
variations of pressure. Infact $I_{\eta}$ is the vertical integral of all the
vertical displacements above the fixed general streamline. The shape
of the obstacle can become very important because the integral of
streamlines vertical displacements depends from it. Moreover to
determine stagnation all the streamlines displacement field
$\eta(x,y,z)$ have to be known. This means that the equations of motion
(Navier-Stokes equations) had to be solved and a non linear solution
of Navier-Stokes equation is not available. Then, to proceed further, a
linear solution of the equation of motion has to be introduced to
evaluate $\eta$ and to insert it into the conservation of energy
obtaining an evaluation of the stagnation point. There is then a formal
contraddiction: when stagnation occurs, streamlines become singular
(e.g., they can split or intersect) then any linear solution is, in
principle, not adapt to describe this behavior. 
To bypass this  difficulty, in this work it is supposed that the
linear solution, jointly with energy conservation, can give some
information useful only to identify the onset of stagnation and not
its behavior. \footnote{It  is possible, using  other technique and
  approximation, taking in account part of the non-linear term and try to give
  some hint on behaviour of streamlines in the stagnation point. This
  aspect is not taking under consideration in this paper and further
  details can be found in \cite{Baines1995b}.}    
Even if this limitation is present, the result seem to be in agreement
  respect the numerical ones; for example  \cite{Smith1989b} has found that, 
  introducing the values of streamlines displacement $\eta(x,y,z)$ obtained by
  his 3-D 
  linear hydrostatic models 
   into  (\ref{eq:U2}), the critical  hill height (i.e. the  minimum  hill
   height  at
   which the flow
   starting to stop) is about 30\% lower than  the
  critical height prediction using the vertical displacement field obtained
  by 
    a  numerical integration of the full non-linear hydrostatic equation using
    in \cite{Smith1989b}.
This fact suggest that the final resolution of the splitting problem
will probably be accomplished using real observational data: this is
not so easy because real data for stagnation position  are difficult to
obtain.
Even though there are evident experimental difficulties,  real data can give some hint on what in reality happens
and what numerical models can be able to see respect analytical ones
and viceversa. 
This paper, following the above idea, will present a comparison of result of analytical models on  the case of Valcanale flood (29 Agust 2003, Valcanale-UD-, Italy).

Finally, It could be argued that if a linear theory is used to provide the field
  of motion, it would be more consistent to use the linearized set  Bernoulli
  equation instead of the complete one (\ref{eq:U2}). This could be done: \cite{Smith1989b}
  found that the critical hill height  is two
  times greater of the values prediction using the complete Bernoulli equation. 
Following \cite {Smith1988}, \cite{Smith1989b} and \cite{Smith1989} we choose to use the exact result  (\ref{eq:U2}).
Thus if any error is present in the derived fields it is because of the
linearization leading to motion equation and not any subsequent linearizations
and it is provides a common method for stagnation diagnosing (analytic) linear and
(numerical) non-linear solution.

{ }


\section{A non-hydrostatic model for the interaction between a flow and a 2-D topography}\label{sec:nohydro}

As shown in the previous section, the non-hydrostatic approch can
become fundamental for obstacles characterized by horizontal scale
comparable with the vertical one. In this case the vertical
acceleration term plays an important role in the flow blocking
phenomena and it has to be taken into account in the evaluation of the
perturbation pressure term of (\ref{eq:bernoulli}). 

To introduce the non-hydrostatic effects a 2-D model is developed and
its governing equations will be solved explicity in the super-critical
(Froude number $\F>1$) and sub-critical ($\F<1$) regime. Then the
streamlines pattern for both these cases and the flow-blocking
dynamics will be described. 

\subsection {Integral representation of solutions}
The development of the model starts from the governing equations of a 2-D incompressible, unviscid and stratified flow in Boussinesq's approximation \cite[]{Emanuel1994b}.
\begin {equation}\label{eq:Boussinesq}
 \left\{ \begin{array}{ll}
 \frac{Du}{Dt}=
 -\frac{1}{\rho_{\infty}}\frac{\p{P^{*}}}{\p{x}}\\
\\
\frac{Dw}{Dt}=-\frac{1}{\rho_{\infty}}\frac{\p{P^{*}}}{\p{z}} -\frac{g\rho'}{\rho_{\infty}}  \\
\\
  \frac{\p{u}}{\p{x}}+\frac{\p{v}}{\p{y}}=0  \\   
 \end{array}
\right.
\end {equation}
where $\rho_{\infty}$ is the unperturbed density field far from the
obstacle. The set (\ref{eq:Boussinesq}) is linearized imposing for the
unperturbed state the horizontal and constant vertical velocity
profile $\boldsymbol{U}=(U_{0},0)$. Moreover the perturbation velocity
field $\boldsymbol{u'}$, using the streamfunction representation, can
be described by 

\begin {equation}\label{eq:streamfunction}
\boldsymbol{u'}=(-\frac{\p{\Psi}}{\p{z}},\frac{\p{\Psi}}{\p{x}} )
\end{equation}

It is then necessary to impose that all the perturbations damp at the
fluid  top  while the linearized impermeability
boundary condition at the ground according to \cite{Baines1995b} is
satisfied, so 
\begin {equation}\label{eq:boundarycondition}
\left\{ \begin{array}{ll}
\Psi=U_{0}h(x)\qquad \textrm{ if }\quad  z=0 \quad\textrm{ ground }\\
\Psi=0        \quad\quad\qquad \textrm{ if }\quad  z=D \quad\textrm{top of fluid }\\
\end{array}
\right.
\end{equation}
In this way the perturbed streamfunction using a linear set of equations satisfys 

\begin{equation}\label{eq:flusso-one}
\left\{ \begin{array}{ll}
{\nabla}^{2}\Psi -
\frac{N^{2}}{g}\frac{\p\Psi}{\p{z}} +
\frac{N^{2}}{{U_0}^{2}}\Psi = 0 \\
\\
\Psi=0 \quad\quad\qquad\quad\textrm{ when }\quad  z=D \\
\\
\Psi=U_{0}h(x)\qquad \textrm{ when }\quad  z=0 \\
\\
\end{array}
\right.
\end{equation}

This set of equations describes the behaviour of a 2-D stratified,
steady, incompressible, parallel flow interacting with a general 2-D
obstacle described by the shape $z=h(x)$. The linearity constraints
the value of the parameters which determine the model. In particular
the constraint is represented by the following two conditions on
Brunt-Vaisala frequency $N$ 
\begin{equation}\label{eq:apprflusso-one}
\left\{ \begin{array}{ll}
\frac{NH}{U_{0}}\ll1\\
\\
\frac{N^{2}D}{g}\ll1\\
\end{array}
\right.
\end{equation}
When the obstacle is represented by a 2-D  hill of
  $H=2000$\m\ high and the flow
is characterized by a upstream velocity $U_{0}=15$\mpers the require startification
frequency that preserves linearity is $N < 7.5 \cdot
10^{-3}$\oneovers. This condition then limits the applicability of the
above developed linear model. In the section ~\S\,\ref{sec:stream} and
~\S\,\ref{sec:stagn} this constrains will be consider.

To proceed further the Fourier trasform in reciprocal space $k$ is
here used to find an integral representation of the solution. Assuming
an orograhy at the ground with this shape
\begin{equation}\label{eq:orography}
z=h(x)= \frac{H}{1+(x/a)^{2}}
\end{equation} 
where $a$ is the half width at half  orography high.  
The solution has the form $\Psi=\Psi_{1}+\Psi_{2}$ where
\begin{equation}\label{eq:solution}
\left\{ \begin{array}{ll}
\Psi_{1}= U_{0}Ha \int_{0}^{c} \cos(kx) e^{-ak} G_{1}dk\\
\\
\Psi_{2}= U_{0}Ha \int_{c}^{+\infty} \cos(kx) e^{-ak} G_{2} dk \\
\\
\end{array}
\right.
\end{equation}
with $ c=N/U_{0}$
\begin{equation}\label{eq:G1}
 G_{1}=\frac{\sin((D-z)\lambda_{1})}{\sin(D\lambda_{1})}
\end{equation}
\begin{equation}\label{eq:G2}
 G_{2}=\frac{\sinh((D-z)\lambda_{2})}{\sinh(D\lambda_{2})}
\end{equation}
and 
\begin{equation}\label{eq:landa1}
\lambda_{1}=\ensuremath{\sqrt{c^{2} - k^{2}}}
\end{equation}
\begin{equation}\label{eq:landa2}
\lambda_{2}=\ensuremath{\sqrt{k^{2}- c^{2}}}
\end{equation}

\subsection{Analytic integration}
 
Obtaining an explicit analytic solution of (\ref{eq:solution}) is not
possible because  the non-hydrostatic term is preserved in the
strating equation  (\ref{eq:Boussinesq}). Moreover it
  can be noticed that, adopting the definition of Froude numeber given
  in \cite{Baines1995b}
\begin{equation}\label{eq:F}
\F=\frac{\pi U_{0}}{N D}
\end{equation}
two different
behaviours exist, according to the different values of the Froude
number.

When $\F>1$ (super-critical regime) the functions $G_{1}$ and $G_{2}$
do not show any singularity on their integral path and the integral of
(\ref{eq:solution}) is defined. For the super-critical regime it
is then possible to adopt a numerical integration of
(\ref{eq:solution}). In particular it is here adopted a ``Monte-Carlo,
importance sampling'' integration with the use of an algorithm for the
generation of pseudo-random numbers. This choice
was taken because for low values of the wave number $k$ and for
low values of $z$, the integral function is nearly an harmonic function
then, an equispatial integration method is not able to reproduce
correctly the orographic profile near to the ground because it
continuously adds harmonic components in phase. The use of a
pseudo-random integration method, in which harmonic components are
added randomly, seems a better approach. A comprehensive tractation of
this integration method can be found in \cite{Gould1996b}. 

In the case $\F<1$ (sub-critical regime) the function $ G_{1}$ admits poles of first order for
\begin{equation}\label{eq:kn}
k_{n}= \ensuremath{\sqrt{c^{2} - (n\frac{\pi}{D})^{2}}} \qquad
\textrm{with}\quad n \in \aleph
\end{equation}
then the integral of (\ref{eq:solution}) exists only in an improper
way called ``Cauchy principal value integral''. The presence of a pole
in the integration path makes impossible the use of numerical
integration methods and the analytic integration of the ``Cauchy
pricinpal value integral''  is needed to deal with the singularity
behaviour but an explicit solution of (\ref{eq:solution}) is not
avaible. 
To overcome this problem in this work a new approach is proposed based on the substitution of $ G_{1} $ and $ G_{2} $ with two new functions  $ \Gone $ and  $ \Gtwo $ such that 

\begin{itemize}
\item The function  $\Gtwo$ has to tend to infinity with the same
  behaviour of  ${G_{2}}$; in this way there is a correct estimate of
  short wave lengths.
\item  The function  $ \Gone $ has to reproduce the same singular
    behaviour and the limit of $\Gone/ G_{1} $ as $k$ tends to the
    singularity has to be unitary.


\item The function  $ \Gone$ and  $\Gtwo$ have to have the same value in $c=N/U_{0}$.
\item The function  $ \Gone$ has to have the same value of ${G_{1}}$
  in $k=0$; this assures that the short wavenumber (i.e., wavelengths
  greater or equal to the horizontal scale of the obstacle) are
  correctly reproduced. 
\item With the new functions $ \Gone$ and $ \Gtwo$ the integral (\ref{eq:solution}) can be solved explicitly.
\end{itemize}

 At this point is necessary to fix some parameter of analytical
  model: with an upstream velocity $U_{0}=15$\mpers, an hill high
  $H=2000$\m\   and a top of  fluid due to the end of Troposphere $D=
  10$\km, linearization conditions (\ref{eq:apprflusso-one}) imposes
  that  the 
  stratification  frequency must be $ N < 7.5 \cdot 10^{-3}$\oneovers. So,
  notice the (\ref{eq:kn}),  there is only one pole $k_{0}$
\begin{equation}\label{eq:ko}
k_{0}=\ensuremath {\sqrt{c^{2} - (\frac{\pi}{D})^{2}} }
\end{equation}
In the following section  ~\S\,\ref{sec:stream} and
~\S\,\ref{sec:stagn} the above value of parameter are considered, so
the function  $G_{1}$  have only one pole in the sub-critical case $(\F
< 1 )$.
 
Following the previously introduced criteria, the new functions that are going to be uses instead of $G_{1}$ and $ G_{2}$ become
\begin{equation}\label{eq:newG2}
\Gtwo=\beta e^{-z(k-c)}
\end{equation}
and 
\begin{equation}\label{eq:newG1}
\Gone= M k +\alpha + W\frac{k(c-k)}{k-k_{0}}
\end{equation}
where
\begin{equation}\label{eq:coeff}
\left\{ \begin{array}{ll}
M=\frac{\beta -\alpha}{c}\\
\\
\alpha=\frac{\sin(c(D-z))}{\sin(Dc)}\\
\\
\beta= 1-z/D\\
\\
\gamma= \sin(\beta \pi)\frac{\pi}{k_{0}^{2} D^{2}} \\
\\
P=\frac{\gamma}{c-k_{0}}\\
\end{array}
\right.
\end{equation}

As we will be shown later on, the wave's dynamic is governed by the parameter $W$; if $W \neq 0$, there is a wave
pattern formation, if $W = 0$ the wave pattern disappears.


Then, inserting the new functions in (\ref{eq:solution}) it is possible to solve explicity the integral \cite[]{Gradshteyn2000b}, obtaining an explicit solution of $ \Psi $ for the sub-critical regime.
It is important to notice that the different behaviour for $\F>1$ and $\F<1$ is a typical example of bifurcation, i.e., differnt physical behaviour as a consequence of an arbitrarily small change in the parameters value.
The case $\F=1$ can not be described by a simply linear theory infact,
for this value of $ F$, equation (\ref{eq:solution}) shows a second
order pole in $ k=0 $ and the integral does not exist, neither in the
sense of ``Cauchy principal value integral''.

\section{The comparison of the streamlines pattern in hydrostatic and non-hydrostatic approaches}\label{sec:stream}

In this section the streamlines pattern of the non-hydrostatic model
are compared with those obtained under the hydrostatic
assumption. This  is done using the 2-D hydrostatic model proposed by
\cite{Baines1995b} making use of the same assumptions 
(steady flow, unviscid, parallel, stratified and incompressible) here adopted.

\begin{equation}\label{eq:hydromodel}
  \Psi=U_{0}h(x)\frac{\sin(c(D-z))}{\sin(cD)}
\end{equation}

From the linearization of boundary condition, fixing as a starting unperturbed level $z=z_{0}$, the streamline is then represented by $z=z_{0}+\Psi/U_{0}$.

All the results presented in the following sections had been obtained
adopting an hill-shaped obstacle whose functional form is
(\ref{eq:orography}) and with the geometrical paramenters $
a=1700$\m\ and $ H=2000$\m.  This value of hill parameter
  corrispond to a
characteristic narrow mountains of Alpine ridge.
It is important to notice that, as it is told before, the topography must be
narrow to have non-hydrostaticity importance.

\subsection{Super-critical regime ($\F>1$)}

\begin{figure*}
\centering
\subfigure[] 
{
 \includegraphics[scale=0.2, angle=270]{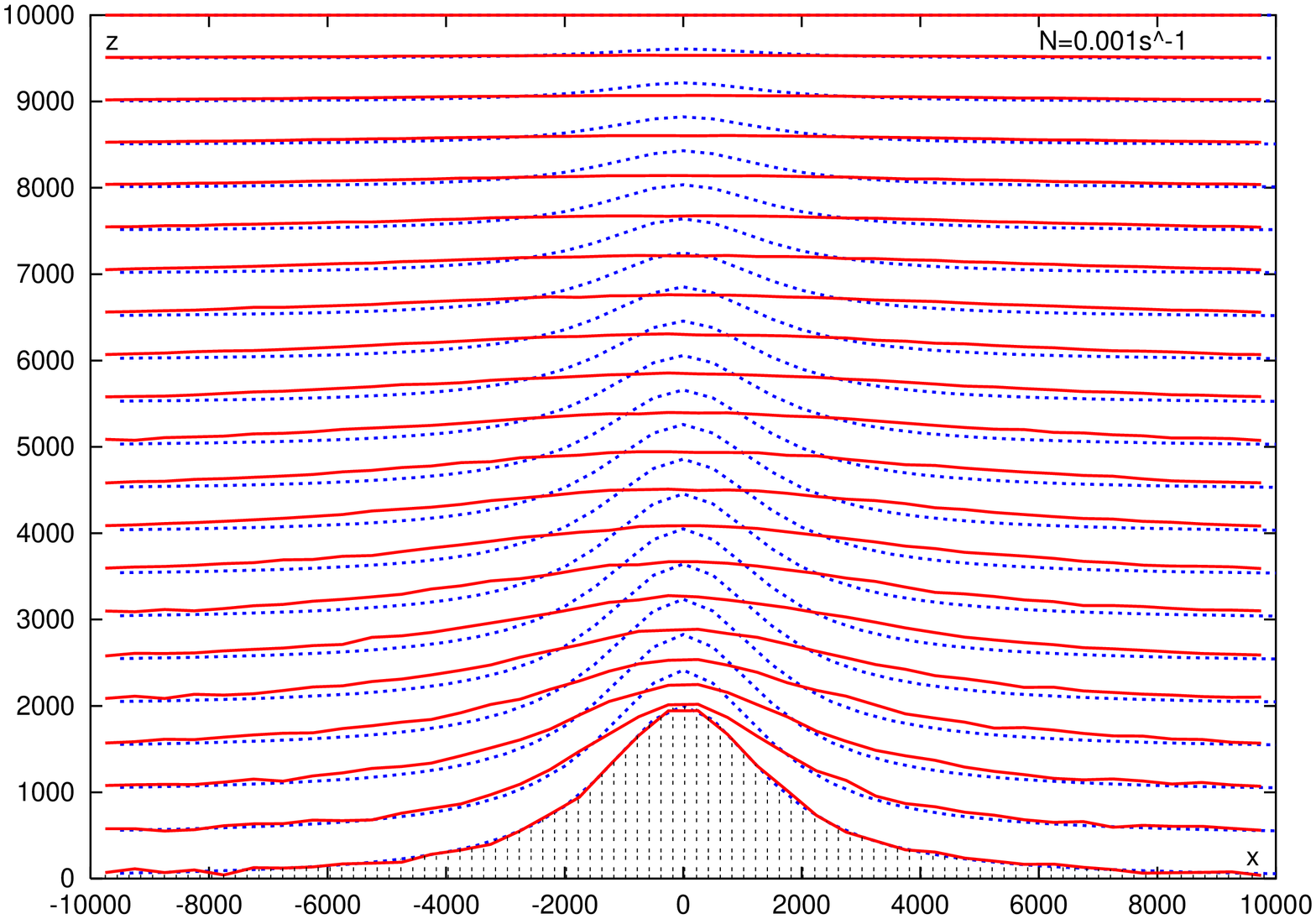}
}
\subfigure[]
{
 \includegraphics[scale=0.2, angle=270]{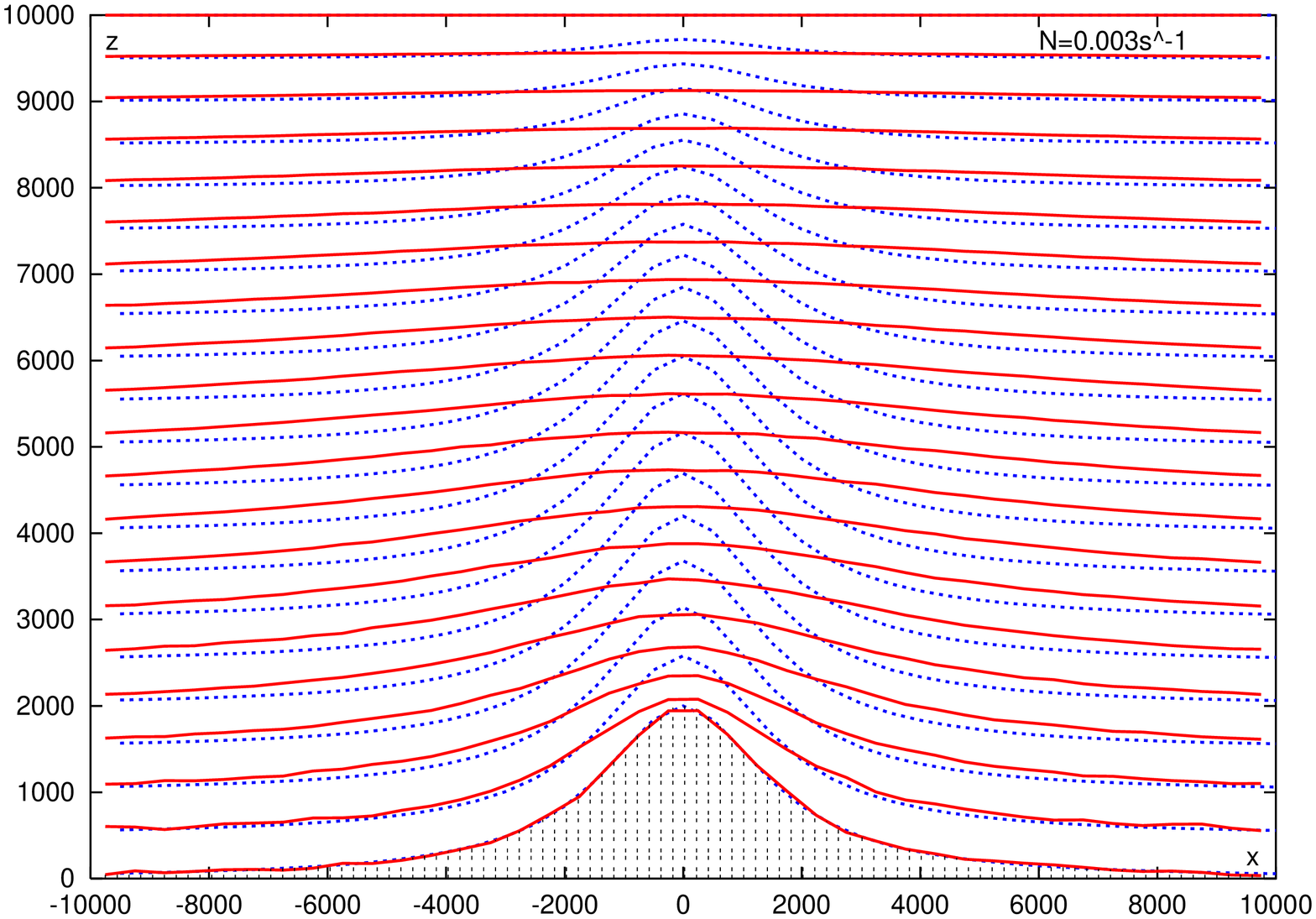}
}
\caption{Super-critical regime of the flow with low
  stratification. Solid lines represent the streamlines of the
  non-hydrostatic model, while dashed lines represent the streamlines
  of the hydrostatic model. Horizontal and vertical axes scale are in
  meters. Panel (\textit{a}) shows the streamlines
  pattern for $N=1\cdot 10^{-3}$\oneovers while panel (\textit{b}) shows the
  results obtained for $N=3\cdot 10^{-3}$\oneovers.}\label{fig:st_super}  
\end{figure*}


In figure \ref{fig:st_super} are displayed the hydrostatic and
non-hydrostatic streamlines patterns. It is clear that the
non-hydrostatic model is characterized by a lower vertical
displacement respect the hydrostatic one at the same stratification
frequency $N$. The origin of this behaviour is in the vertical
acceleration term that the hydrostatic model can not take into
account.
Infact the increase of vertical streamlines displacement and velocity,
due to the  continuity equation as a consequence of  narrowing of
troposphere, gives rise to the formation of a low pressure zone at
hill's top.  This low pressure zone constrains the vertical
streamlines displacement because the fluid in the upper part of
troposphere  is pushed downward by this pressure deficit. 
This lower pressure is more accentuated in the
non-hydrostatic model because the extra term represented by the
velocity vertical variation. 
Then streamlines displacement is reduced
in the non-hydrostatic model by the action of this counteracting
pressure gradient. 



\begin{figure*}
\centering
\subfigure[]
{
 \includegraphics[scale=0.4, angle=0]{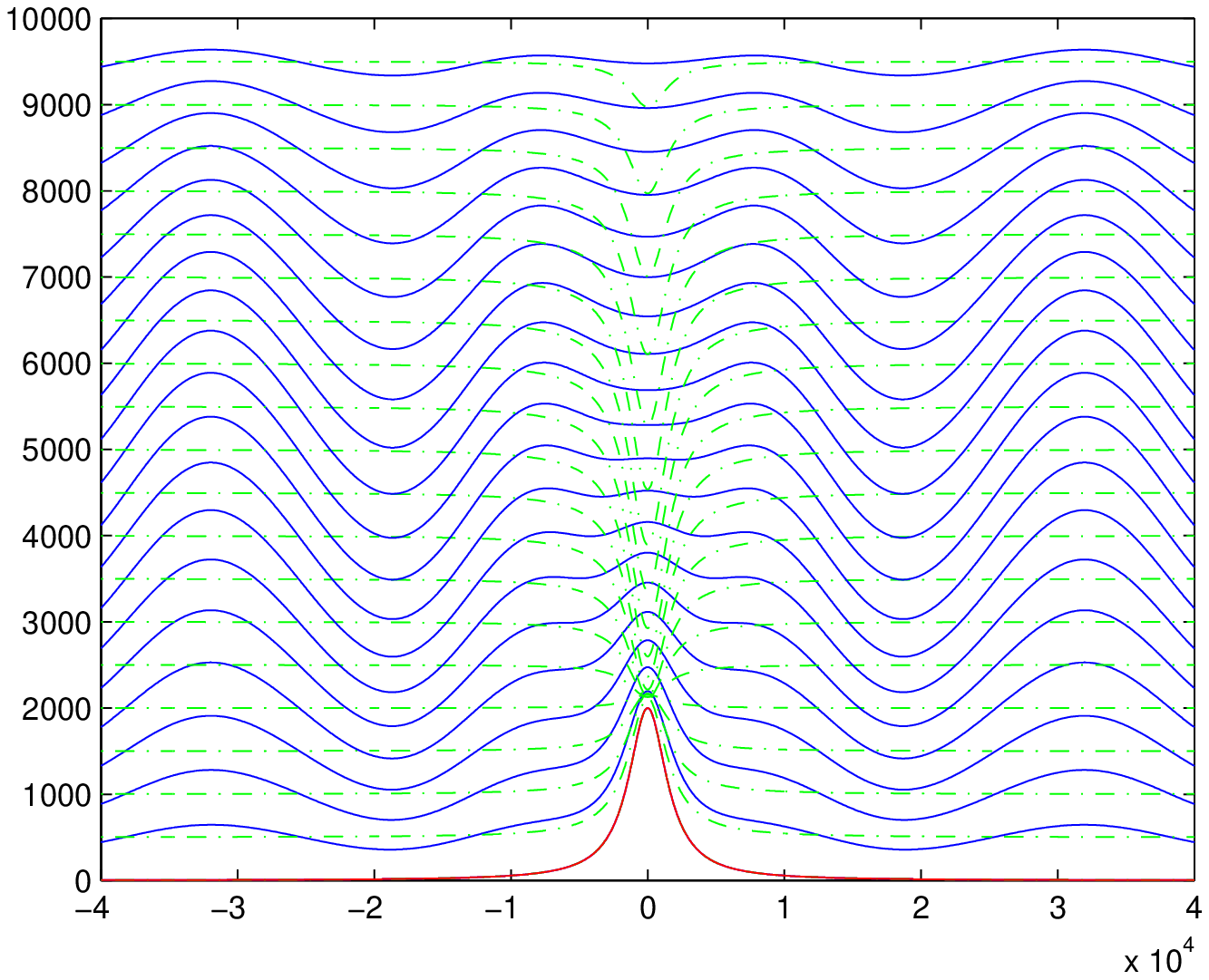} 
}
\subfigure[]
{
 \includegraphics[scale=0.4, angle=0]{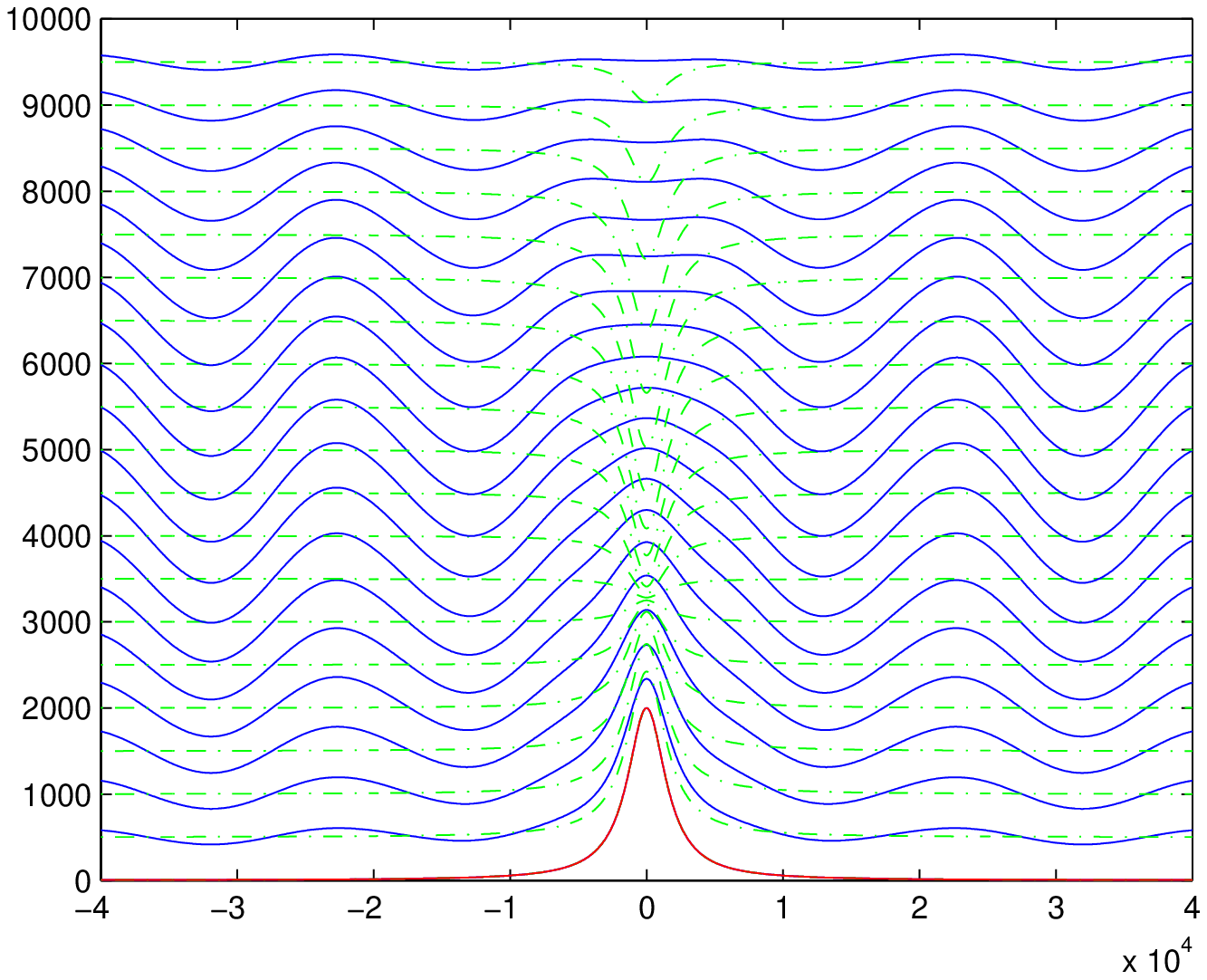}
}
\caption{Sub-critical regime of the flow. Solid lines represent the
  streamlines for the non-hydrostatic model while dashed lines
  represent the streamlines for the hydrostatic model. Horizontal and vertical axes scale are in
  meters. Panel
  (\textit{a}) shows the streamlines pattern for $N=6\cdot
  10^{-3}$\oneovers
  while panel (\textit{b}) shows the streamlines pattern for
  $N=7\cdot 10^{-3}$\oneovers.}\label{fig:st_sub_waves} 
\end{figure*}


Before to conclude it has to be noted that the vertical displacement
increases with the increasing of $N$. This behaviour can be observed
both in the hydrostatic and non-hydrostatic model and  might seem
counter-intuitive (one could think that when stratification increases
then the restoring forces become larger). This behavior is explained
considering that the inertia of the lower-levels parcels increases
more than the inertia of the upper-levels parcels when stratification
increases. 
So, even if the
  restoring force becomes larger   as stratification increase,  the
   difference of masses of different air level parcel is crucial for
  the developing of perturbation pressure that play an important rule
   in the streamlines formation and, as it will be possible to see
  later,  in flow-blocking dynamics.

\subsection{Sub-critical regime ($\F<1$)}

\begin{figure*}
\centering
\subfigure[]
{
 \includegraphics[scale=0.4, angle=0]{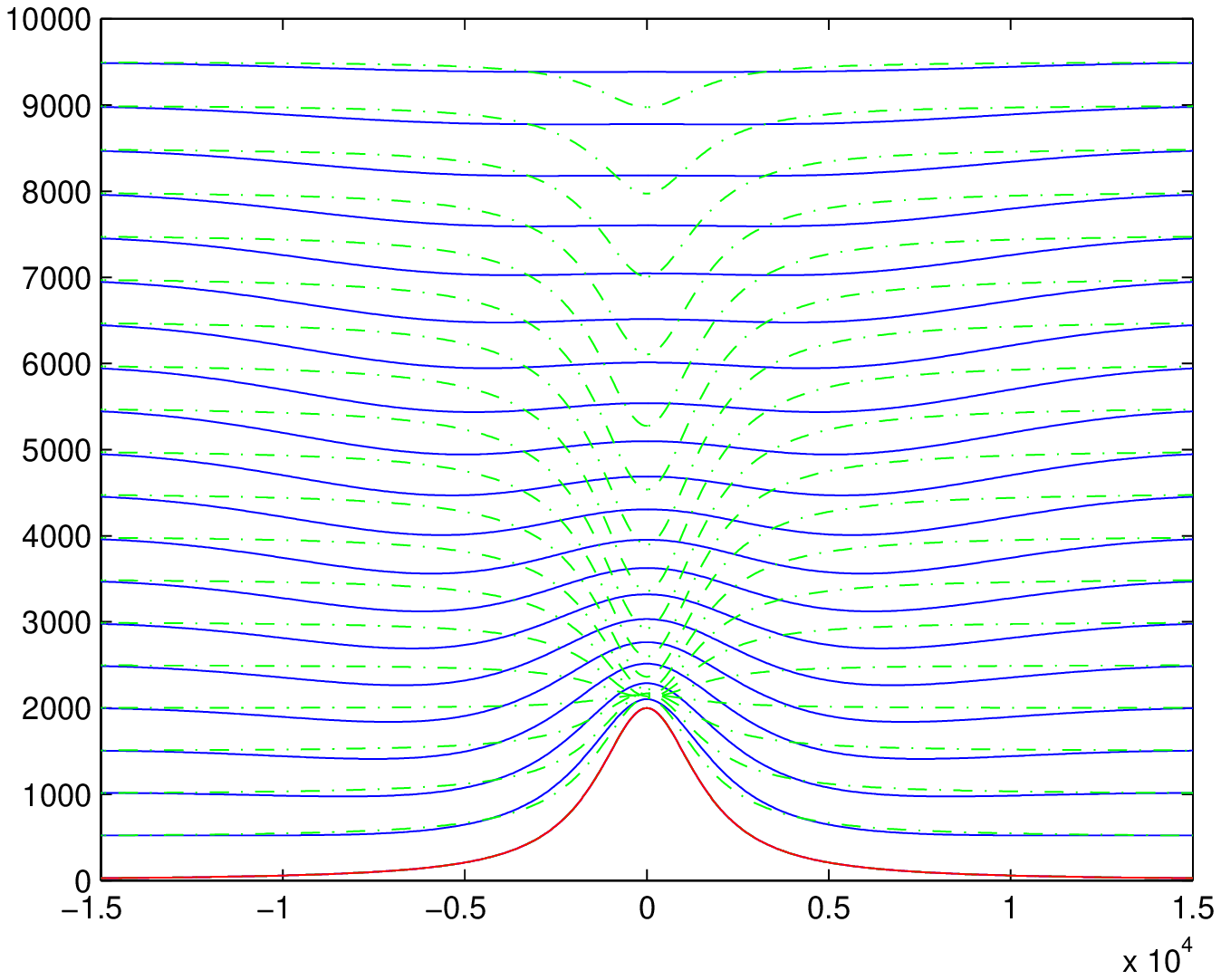} 
}
\subfigure[]
{
 \includegraphics[scale=0.4, angle=0]{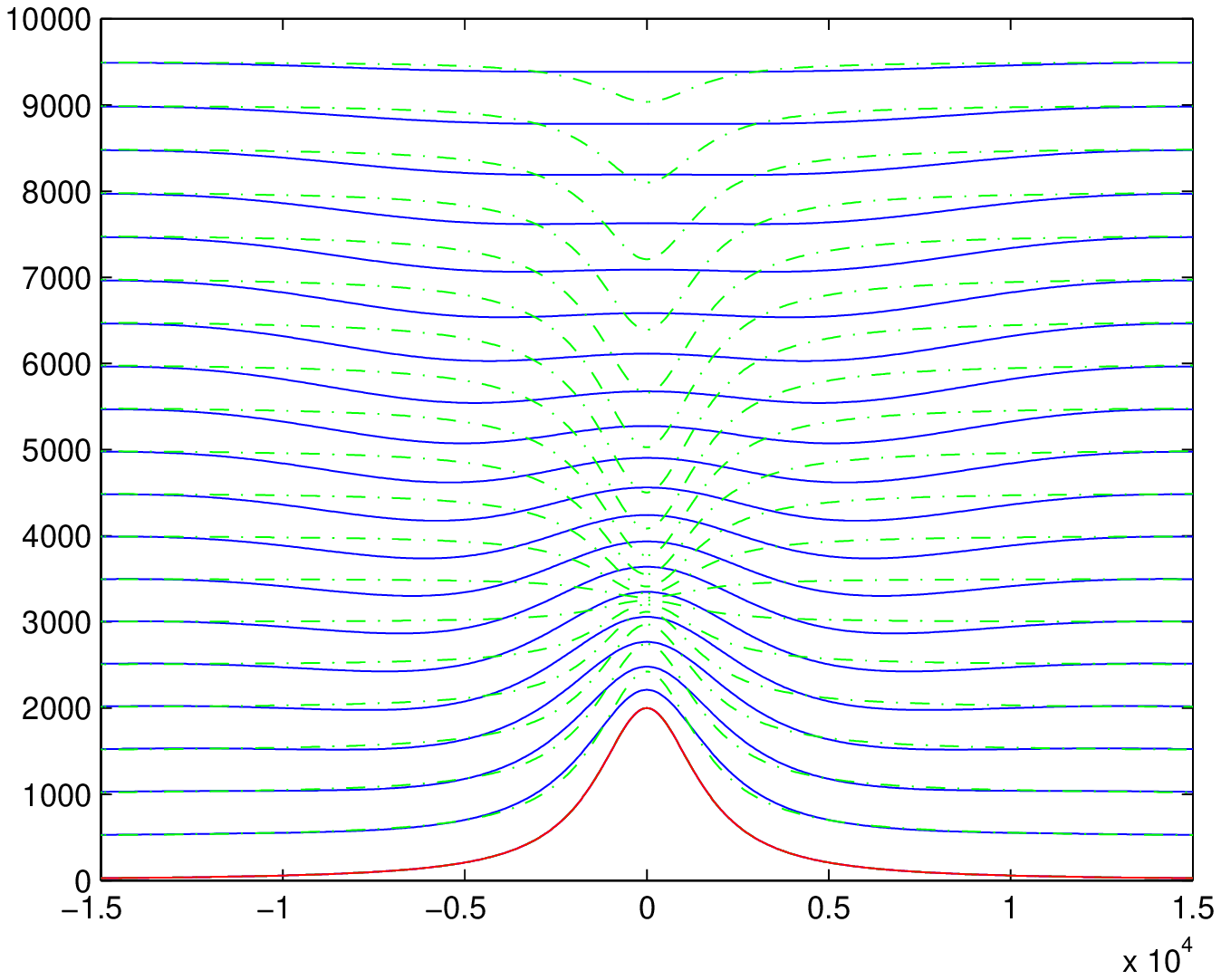}
}
\caption{Sub-critical regime of the flow with high
  stratification. Solid lines represent the streamlines of the
  non-hydrostatic model while dashed lines represents the streamlines
  for the hydrostatic model.Horizontal and vertical axes scale are in
  meters. Panel (\textit{a}) shows the streamlines
  pattern for $N=6\cdot 10^{-3}$\oneovers while panel (\textit{b})
  shows the streamlines pattern for $N=7\cdot
  10^{-3}$\oneovers.}\label{fig:st_sub_nowaves}  
\end{figure*}


Considering all the terms present in (\ref{eq:newG1}) the results
shown in figure \ref{fig:st_sub_waves} are obtained. In that picture
it is clearly recognizable the formation of stationary waves. These
results are very similar to those found in \cite{Keller1994}.  The
formation of these waves is characteristic of the singularity, infact
for a wind velocity of $ 15$\mpers and a  stratification frequency
of the order of $ N= 7 \cdot 10^{-3}$\oneovers there is a pole at  
\begin{equation}\label{eq:ko}
k_{0}= \ensuremath{\sqrt{c^{2} - (\frac{\pi}{D})^{2}} }= 3.45 \cdot 10^{-4} \oneoverm
\end{equation}
which corresponds to a wavelength of $ \lambda \approx 18$\km that is clearly recognizable in figure \ref{fig:st_sub_waves}.

Before to proceed further it has to be noted that the streamline waves
pattern is symmetric to the top of the hill while from real cases it
is clear that the waves pattern can be present only downstream to the
hill. This is a spurious effect of the linearization,  infact the
(\ref{eq:flusso-one}) does not change for the reflection of the
variable $x$. To avoid this spurious effect it is sufficient to fix
$W=0$ in (\ref{eq:newG1}) and the streamlines pattern that one obtain
is shown in figure \ref{fig:st_sub_nowaves}. 
It is then interesting to notice that even if the waves are produced
by the obstacle, their wavelength has not the same order of magnitude
of the hill's horizontal scale. Infact, being triggered by the
singular point $k_{0}$, the waves depend only from the stratification
and from the velocity profile. 

In figure \ref{fig:st_sub_nowaves} it is possible to clearly recognize the
velocity intensification at the top of the obstacle represented by the
streamlines concentration in that region. This intensification is a
well known effect to mountain-hickers and it is due to the falling
down of the upper-level streamlines which is not present in the
super-critical stratification.  
This different behaviour comes out because  in the sub-critical regime
the upper-level parcels fall down in a denser environment that gives
them the necessary up-ward lifting force to return to their initial
level when they return far from obstacle. The obtained stationary
pattern is represented in figure \ref{fig:st_sub_nowaves}. This
situation can not take place in the super-critical regime, when a
downward displacement similar to that of the sub-critical case, cannot
receive the same lifting force because of a less stratified
environment. 

It is interesting to notice that the non-hydrostatic model gives a
lower intensification of wind speed at the top of the hill respect the
hydrostatic ones at the same stratification frequency.
A possible description of this effects can be explained,  as was done
in the super-critical case, considering the Bernoulli equation even if,  in
sub-critical case,  the streamlines pattern is more complicated respect the
super-critical one. 
In this case the acceleration of the fluid and the increase of
vertical displacement of lower level streamlines gives the formation
of a low pressure area at hill's top while 
the falling of upper  streamlines 
could be enough to compensate the fluid acceleration and so this falling
streamlines could give an increase of pressure: infact the
mid-level streamlines, after an initial falling, seem to moves
upward due to the acting of this increase of pressure as it is
possible to see in figure \ref{fig:st_sub_nowaves}.
So, this more complicated pressure pattern gives rise to a lower
intensification of wind speed at the top of the hill and, as it will
be noted later, is the responsable of flow-blocking in sub-critical flow.

Finally it is worth to be noticed that,  also in this case, the
vertical displacement increases with the increasing of
stratification (the altitude of the wind speed intensification  increase grows with
$N$). The explanation given again makes use of the increasing of
inertia with the increasing of stratification. 
Lower-level parcels
moving upward find themselves immersed in a layer of fluid less dense,
then with a low capability of counteracting with the buoyancy force
their upward motion.

Before to conclude this section it is again important to state that, both in the super-critical and sub-critical regime, the non-hydrostatic effect consists in a smoothing of the streamlines geometry and the smoothing effect is concetreted near to  the top of the obstacle.

\section{Non-hydrostatic effects on flow-blocking}\label{sec:stagn}

Knowing the streamfunction for the non-hydrostatic case, it is
possible to can calculate the non-hydrostatic term $\Gamma_{nh}$ of
(\ref{eq:gammahydro}) due to the vertical acceleration and use it in
(\ref{eq:U2}). Four cases for different stratification frequencies
$N=1,3,6,7 \cdot 10^{-3}$\oneovers are taken into account. The first two
frequencies (i.e., $N=1,3 \cdot 10^{-3}$\oneovers) correspond to the
super-critical regime while the last two (i.e., $N=6,7 \cdot
10^{-3}$\oneovers) to the sub-critical regime. 

\subsection{Hydrostatic result}

\begin{figure}
\begin{center}
  \includegraphics[scale=0.3, angle=270]{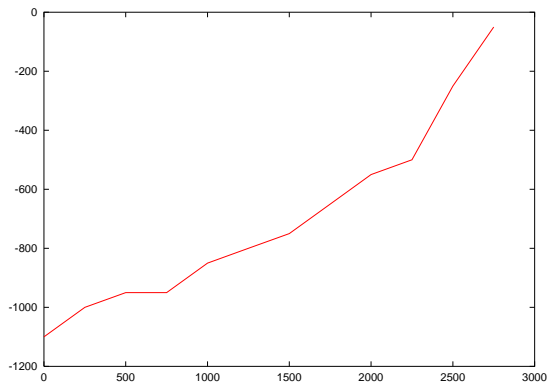} 
\caption{Flow-blocking and hydrostatic approximation. The unperturbed
  streamine level is reported in the horizontal axis, in meters, while the
  abscissa of stagnation (i.e, the distance from the top of the hill
  where the flow-blocking occurrs) is reported, in meters, in the vertical
  axis. This picture is obtained with a stratification $ N=3 \cdot
  10^{-3}$\oneovers.}\label{fig:stagn_hydro} 
\end{center} 
\end{figure}

In the equation (\ref{eq:U2}) the vertical advection term
(\ref{eq:gammahydro}) is null because a hydrostatic model is here
used, then the streamlines displacement is given by
(\ref{eq:hydromodel}). The relationship between the unperturbed
streamline level and stagnation abscissa is shown in figure
\ref{fig:stagn_hydro}. The only stratification frequency that admits
stagnation is $N= 3\cdot 10^{-3}$\oneovers, which corresponds to the
streamlines pattern where the vertical displacement larger, as can be
seen in figure \ref{fig:st_super}. 
The
explanation of this fact can be found in the behaviour of streamline
pattern. 
%
In a super-critical hydrostatic flow,  stagnation  occours when the
streamlines does not have a sufficient kinetic energy  to
trasform into potential to produce the vertical displacement
request. 
On the contrary, the sub-critical flow blocking, as it will be remark
in the next section, happens when the falling of upper streamlines
creates a pressure configuration at the top of hill that
could stop low-level streamlines

A last comment can be done on stagnation abscissa: there are
streamlines, in the hydrostaic case, that
stopped in the proximity of hill's top.


\subsection{Non-Hydrostatic result}

\begin{figure*}
\centering
\subfigure[]
{
 \includegraphics[scale=0.2, angle=270]{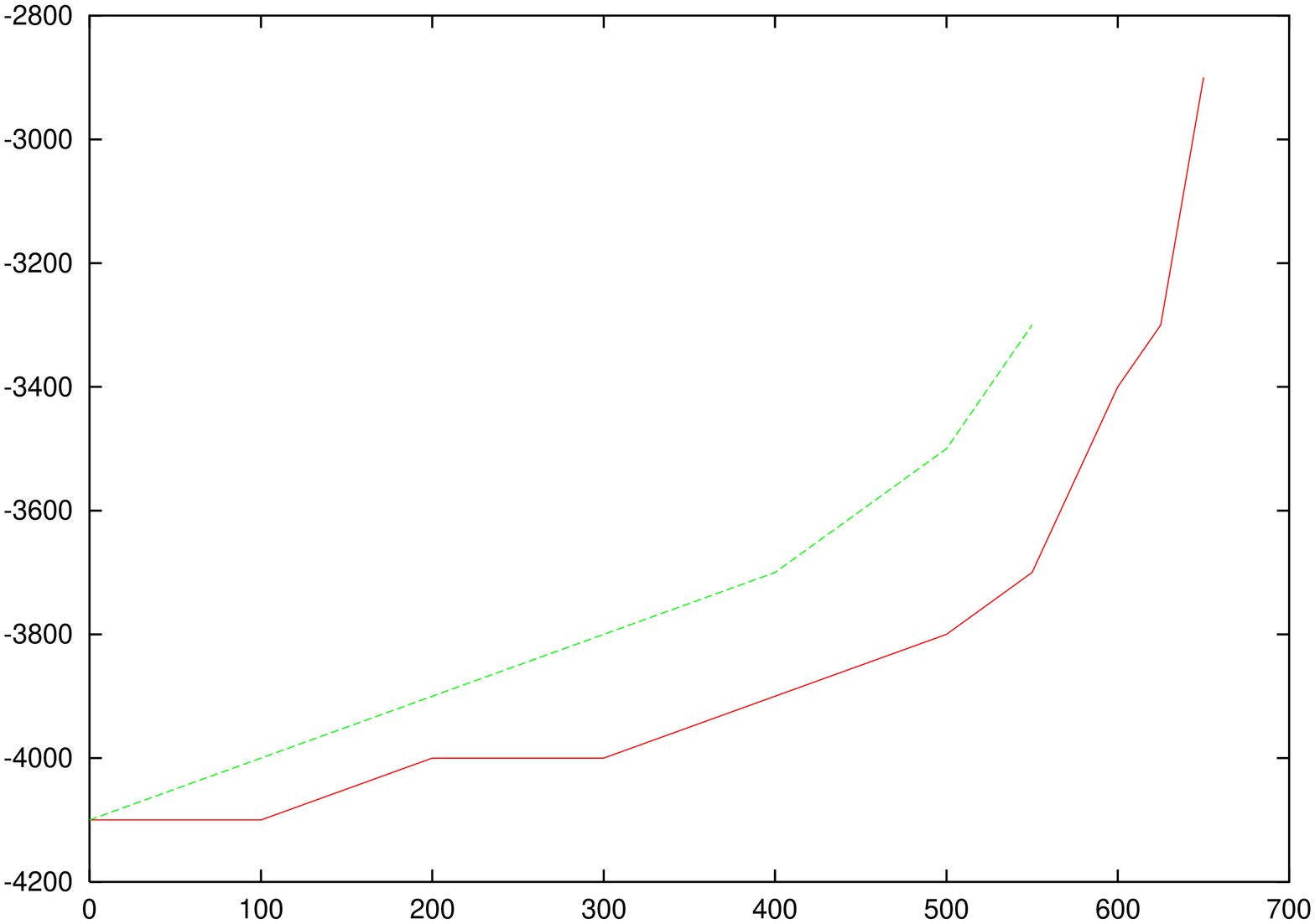} 
}
\subfigure[]
{
 \includegraphics[scale=0.2, angle=270]{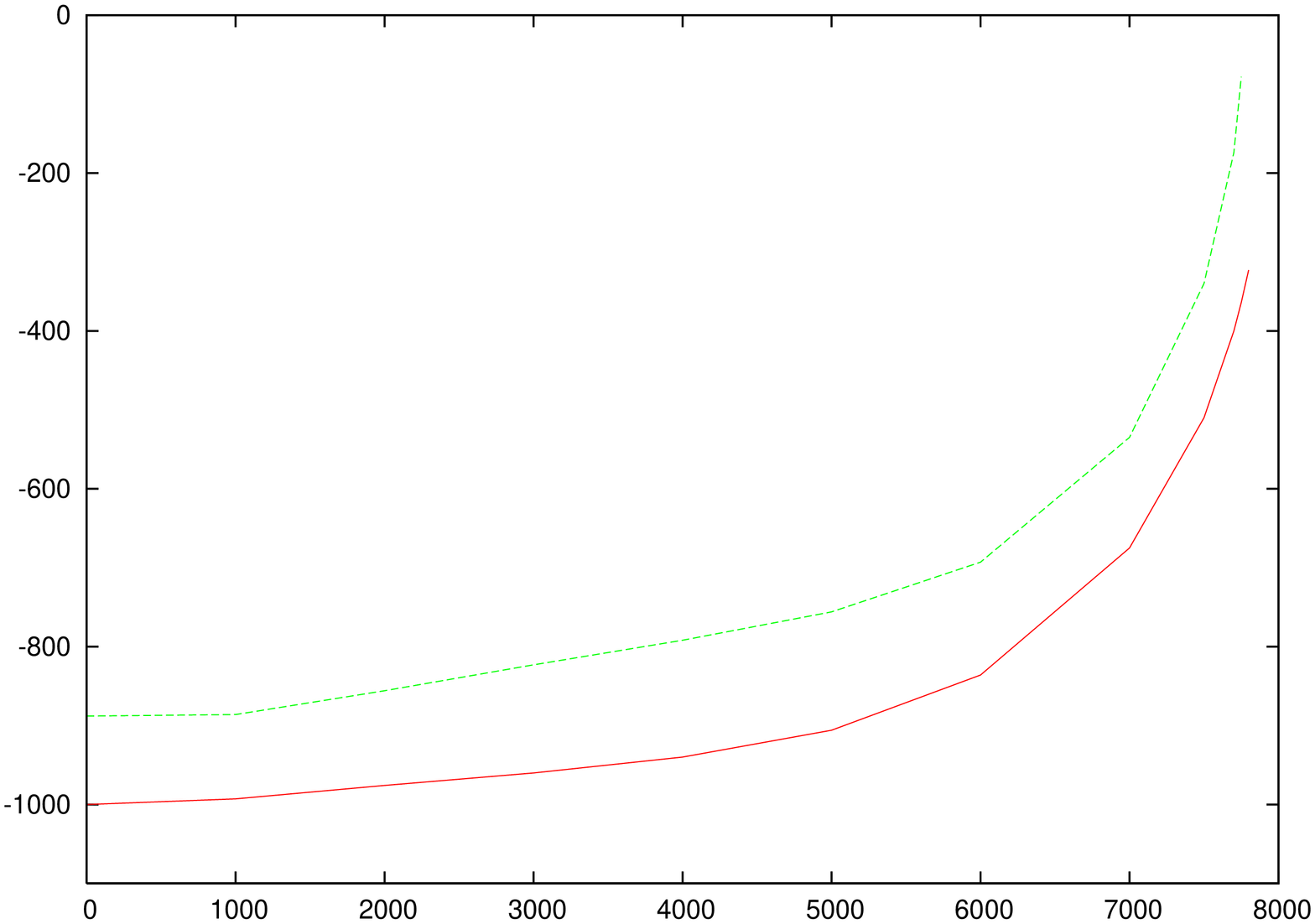}
}
 \caption{Non-hydrostatic case. In the horizontal
  axis it is reported the unpertubated streamline level (in meters) while the
  vertical axis reports, in meters, the abscissa of stagnation, i.e. the distance
  from the top of the hill where the fluid
  stops. Panel (\textit{a}) reports the
  super-critical regime. Dash upper  lines correspond to the stratification
  $N=3 \cdot 10^{-3}$\oneovers while solid lower lines correspond to $N=1 \cdot
  10^{-3}$\oneovers. Panel (\textit{b}) reports the sub-critical regime.
  Dash upper lines correspond to the stagnation $N=7 \cdot 10^{-3}$\oneovers
  while solid lower lines to $N=6 \cdot 10^{-3}$\oneovers.}\label{fig:stagn_nohydro}  
\end{figure*}


In the non-hydrostatic case flow stagnation can take place for all the
considered stratification frequencies. 

In the super-critical regime the behaviour of
the abscissa of stagnation is shown in the panel $a$ of figure
\ref{fig:stagn_nohydro}. This behaviour is similar to that evidenced
in the hydrostatic regime even if in the non-hydrostatic case the
abscissa of stagnation is upward limited and does not pass the abscissa
of $2500$\m\ from the top of the hill. The explanation of this fact can
be found in the role of the streamlines curvature. In figure
\ref{fig:stagn_nohydro}a it is possible to observe that as all the
stagnation abscissas are positionated before of the streamlines'
flexums (for the ground streamline the flexum is at $a \approx 1700$\km, value that corresponds to the flexum of
(\ref{eq:orography})). This effect is similar to that experienced by a
driver when he or she is running on the positive curvature of a road
and his/her car is forced downward while on the negative curvature of
the road the car is lifted. The same effect takes place in fluids but
only a non-hydrostatic model can keep into account this mechanism
because it is the only model that can consider vertical
accelerations. 

In the sub-critical regime the situation is quite different. The
inertia of the lower-levels fluid parcels pushes the fluid beyond the
flexum and stagnation occurs later than in the super-critical regime.
The other important things to be noticed is that
large portions of the fluid are blocked in the sub-critical regime
(nearly the $8$\km of fluid nearer to the ground). This situation
corresponds to the pattern in figure \ref{fig:st_sub_nowaves}  where the upper
streamlines fall toward the obstacle. This means that large amounts of
kinetic energy become potentially available. This fact is in agreement
with the findigs of \cite{Schar1993b}, \cite{Smolarkiewcz1989} and,
\cite{Castro1983} where stagnation is associated with the vorticity
generation. 

Concerning the role of stratification frequency on stagnation in
figure \ref{fig:stagn_nohydro} it can be shown that when $N$
increases, stagnation occours later, i.e., for smaller abscissas. This
fact is in agreement with the streamlines pattern, in fact when
stratification increases the vertical displacement is larger. This
result  might seem countrintuitive but it can be explained taking into
account the fact that when stratification increases, restoring forces
increase but the inertia of the lower parts of the fluid increases as
well. This inertia is strictly connected to the generation of pressure
perturbation,  that play a crucial roll in flow-blocking dynamics, and a non-hydrostatic model can better keep into account
this fact in comparison with hydrostatic models. Even if this result 
is obtained with a linear model valid only for a restrict range of
stratification frequencies it can give a useful hint toward the
interpretaion of real cases, in particular of what happend in
Valcanale (UD), Italy, during the $ 29^{th} $ August 2003. 
\begin{figure}
\begin{center}
  \includegraphics[width=0.45\textwidth, angle=0]{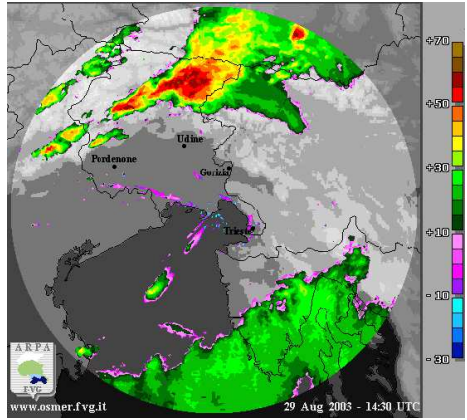} 
\caption{Radar image: a moist and stable stratified flow with $ N=0.03$\oneovers
  interact with 2-D ridge. The flow overtaking the first ridge but stopped on
  the second (red and dark region) where starting convection and an intense
  orographic rains. Photo  made by doppler radar station of Fossalon (Ud),
  Italy by Meteorological regional service O.S.M.E.R.}\label{fig:radar}
\end{center} 
\end{figure}

In that day a flow strongly stratified in the lower levels ($ N=3
\cdot 10^{-2}$\oneovers compared to the mean value of $ N=7 \cdot
10^{-3}$\oneovers ) moving from south and interacting with the
orographic ridge (Julian Preals and Alps) overcame the first ridge
producing large amount of rain (nearly $400 mm$ in four hours) and two
casualties only on the further inner ridge (Julian Alps) where
convection took place, as can be seen from the radar image shown in
figure \ref{fig:radar}.

\section{Conclusions}

This paper presents a study on the interactions beetween stratified flows and orography carried out developing an analytical model. In particular, starting from the previous works of \cite{Smith1989} and \cite{Keller1994}, here the non-hydrostatic terms are kept into account in a simplified 2-D model to evaluate their effects on streamlines patters and flow-blocking.
Following this idea an integral solution of the 2-D non-hydrostatic model had been found, whose behaviour is described by way of the Monte-Carlo sampling integration method for the super-critical regime and by way of a newly developed explicit integration approch in the sub-critical regime.

The main results of this work can be summarized in the following points:

\begin{itemize}
\item The non-hydrostatic effects are important for a topography
  characterized by a horizontal scale comparable with the vertical
  scale. In this case, in fact, the change in the topography
  curvature,  then in the streamlines curvature that is connected to
  perturbative pressure, becomes important.

\item The non-hydrostatic effects produces a general smoothing of the
  streamlines pattern. These effects  can be explained taking into
  account the presence of the high/low pressure area, due to the
  complexity of streamlines picture, that a non-hydrostatic model are
  much able to see. 


 
\item The streamlines pattern shows, for the sub-critical regime, the formation of waves downstream to the topography, that hydrostatic models can not reproduce. The wavelength of this undulatory pattern is connected to the properties of a singular point in the integral function which depends only from the stratification frequency and from the upstream velocity of the flow. The intensification of flow velocity at the topography top is observed in the sub-critical regime. 

\item Stagnation occurs upstream to the topography and hydrostatic and
  non-hydrostatic models show a very different behavior in
  flow-blocking. In hydrostatic models stagnation occurs on the top of
  topography and only for large vertical displacements of streamlines
  (i.e, in the super-critical regime). 

\item In the non-hydrostatic model, for the super-critical regime,
  only lower-levels streamlines are blocked and stagnation does not
  occur beyond the flexum abscissa. This behaviour can be explained
  taking into account the role of streamline's curvature which is
  connected to the vertical perturbation pressure (lifting in the
  negative curvature part of streamlines and downward restoring force
  in the positive curvature part of streamlines) that can not be
  observed in hydrostatic models. 

\item  In the non-hydrostatic model, for the sub-critical regime, the
  stagnation is due by a formation of complex pressure pattern at the
  top of hill. 
  Moreover stagnation in sub-critical case occurs beyond the flexum's abscissa upstream to
  topography. This because the inertia of the lower-levels parcels is
  larger than the inertia of the upper-levels parcels. In the
  sub-critical regime the amount of fluid that is blocked is very
  large: this means that a large amount of kinetic energy is made
  avaible. This source of kinetic energy might become a source of
  vorticity production connected with the flow-blocking as  previously
  observed by \cite{Schar1993b}, \cite{Smolarkiewcz1989} and
  \cite{Castro1983}. 

\item As stratification increases, the flow stops later than a less
  stratified flow.
For this reason stratification
  parameters alone are not sufficient to detect the flow-blocking
  situation because stratification and curvature effects are linked to
  the perturbation pressure that can have an important role on
  flow-blocking, as shown in this work.  
So this work suggest the opportunity to connect
  perturbation pressure to stratification and curvature
  parameters. 

\item This work suggest the possibility to construct a faster
  evaluation method of  stagnation phenomena:
  the result of a linear
model, that can be run faster and for a more smaller spatial mesh,
could be use to describe the 
flow-blocking onset. This could be an important application for the
wheater forecasting where the evaluation time is an important
parameter to minimize in many circumstances. 

\end{itemize}

\begin{acknowledgments}
The authors aknowledge the FORALPS Project (Interreg IIIB - Alpine Space) in the frame of which this work was partially carried out.
\end{acknowledgments}


\bibliographystyle{jfm}
\bibliography{Gladich_arxiv}

\end{document}